\documentclass[12pt,a4paper]{article}
\usepackage{graphicx,color,amsmath, amssymb,cite,multirow}
\usepackage[english]{babel}
\usepackage{t1enc}
\usepackage[latin2]{inputenc}
\begin{document}
\textwidth=135mm
 \textheight=200mm

\begin{center}
{\bfseries Femtoscopic results in Au+Au and p+p  from PHENIX at RHIC
\footnote{{\small Talk at the VI Workshop on Particle Correlations and Femtoscopy, Kiev, September 14-18, 2010.}}}
\vskip 5mm
M. Csan\'ad$^\dag$ for the PHENIX Collaboration
\vskip 5mm
{\small {\it $^\dag$ Eötvös University, Department of Atomic Physics, Pázmány P. s. 1/A, H-1117 Budapest, Hungary}}
\\
\end{center}
\vskip 5mm
\centerline{\bf Abstract}
Ultra-relativistic gold-gold and proton-proton collisions are investigated in
the experiments of the Relativistic Heavy Ion Collider (RHIC). In the last
several years large amount of results were revealed about the matter
created in these collisions. The latest PHENIX results for femtoscopy and correlations
are reviewed in this paper. Bose-Einstein correlations of charged kaons in 200 GeV Au+Au
collisions and of charged pions in 200 GeV p+p collisions are shown. They are
both compatible with previous measurements of charged pions in gold-gold
collisions, with respect to transverse mass or number of participants scaling.

\vskip 10mm

\section{Introduction}
Ultra-relativistic collisions of Au nuclei are observed at the experiments
of the Relativistic Heavy Ion Collider (RHIC) of the Brookhaven National Laboratory,
New York. The aim of these experiments is to create new forms of matter that
existed in Nature a few microseconds after the Big Bang, the
creation of our Universe.

A consistent picture emerged after the first three years of
running the RHIC experiment: the created hot matter acts like a
liquid~\cite{Adcox:2004mh}, not like an ideal gas some had
anticipated when defining the term QGP.
The nuclear modification factor is ratio of yield in Au+Au collisions
over the yield in p+p collisions, scaled by the number of binary
nucleus-nucleus collisions in a Au+Au collision. It has been measured for several
hadron species at highest $p_t$, most recently
$\eta$ and $\phi$ mesons\cite{Adare:2010dc}. This
confirms the evidence for a dense and
strongly interacting matter. Direct photon
measurements, which require tight control of experimental
systematics over several orders of magnitude, show that high
$p_t$ photons in Au+Au collisions are not
suppressed~\cite{Adler:2005ig}. This observation makes
definitive the conclusion that the suppression of high-$p_t$
hadron production in Au+Au collisions is a final-state effect.

A very important tool to understand the geometry of the matter created
at RHIC is that of Bose-Einstein correlations, or otherwise
interferometry of bosons. In present
proceedings paper we do not detail the theory,
simply refer to ref.~\cite{Csorgo:1999sj}. In the next
sections we will detail recent measurements about
pion and kaon interferometry.

\section{Kaon interferometry in Au+Au collisions}
The observations of extended, non-Gaussian, source size from two-pion
correlations~\cite{Adler:2006as} make the measurement of two-kaon correlations
important for understanding the contribution from decays of long-lived resonances. 

This analysis is described in detail in ref.~\cite{Afanasiev:2009ii}.
PHENIX used $\sim$~600 million minimum bias events, triggered by the coincidence
of the Beam-Beam Counters (BBC) and Zero-Degree Calorimeters (ZDC) with collision
vertex $|z|<30$~cm. Charged kaons were tracked and identified using the drift chamber 
(DC), pad chambers (PC1,PC3) and PbSc Electromagnetic Calorimeters 
(EMCal) to cover pseudorapidity $|\eta|<0.35$ and azimuthal 
angle $\Delta\phi=3\pi/4$. Momentum resolution in this case was 
$\delta p/p \simeq 0.7\% \oplus 1.0\%\times p$ (GeV/$c$). 
Backgrounds were reduced by requiring 2 $\sigma$ position match 
between track projections and EMCal hits, and 3 $\sigma$ match for 
PC3. Until a transverse momentum of $\sim$0.9 GeV/$c$ kaons
and pions can be separated via timing information. Above
that limit PID cuts have to be introduced, the selection
in this case was that we identify particles as kaons if they
are within 2 $\sigma$ of the theoretical mass-squared of kaons
and are at least 2 $\sigma$ away from the pion and also the
proton mass. With this, the contamination level is $\sim$4$\%$
from pions, and $\sim$1$\%$ from protons at $p_t\sim1.5$ GeV/$c$.

\begin{figure}
\centering
\includegraphics[width=0.73\linewidth]{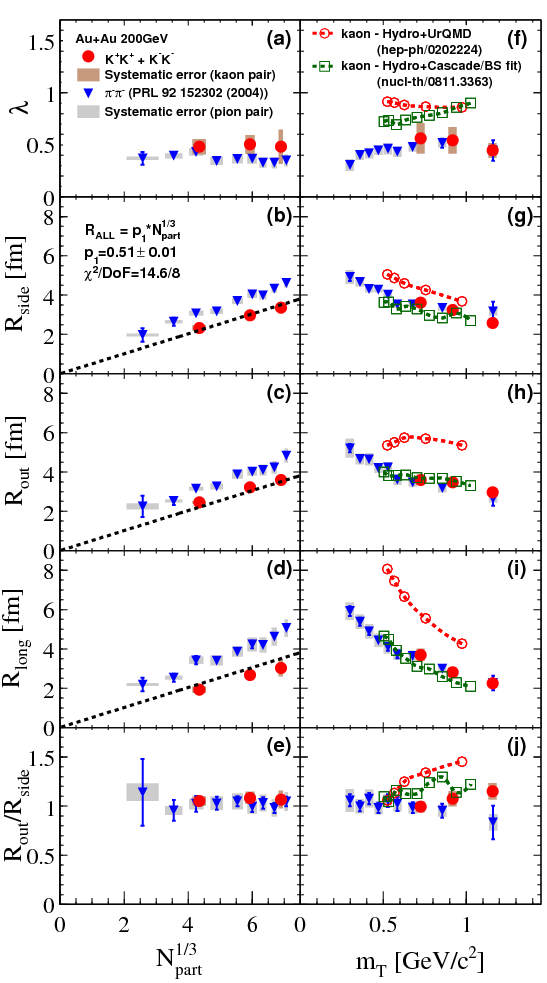}
\caption{3D Gaussian HBT radius parameters
for charged kaon pairs are plotted as a function of $N_{\rm part}^{1/3}$ (left),
and as a function of $m_t$ (right).\label{f:kaonradii}}
\end{figure}

We find that the number of participants ($N_{part}^{1/3}$) dependence
of 3D correlation radii is linear as shown on fig.~\ref{f:kaonradii}.
The transverse momentum ($m_t$) dependence of these radii follows
the same scaling as in case of pions, predicted from hydro models,
see for example~\cite{Csanad:2008gt}. In case of imaging,
a non-Gaussian tail is revealed for radii greater than 10~fm.
This suggests that earlier finding of large tails of pion
imaged source functions are not due to resonance decays but
show truly enlarged sources.

\begin{figure}
\centering
\includegraphics[width=1.0\linewidth]{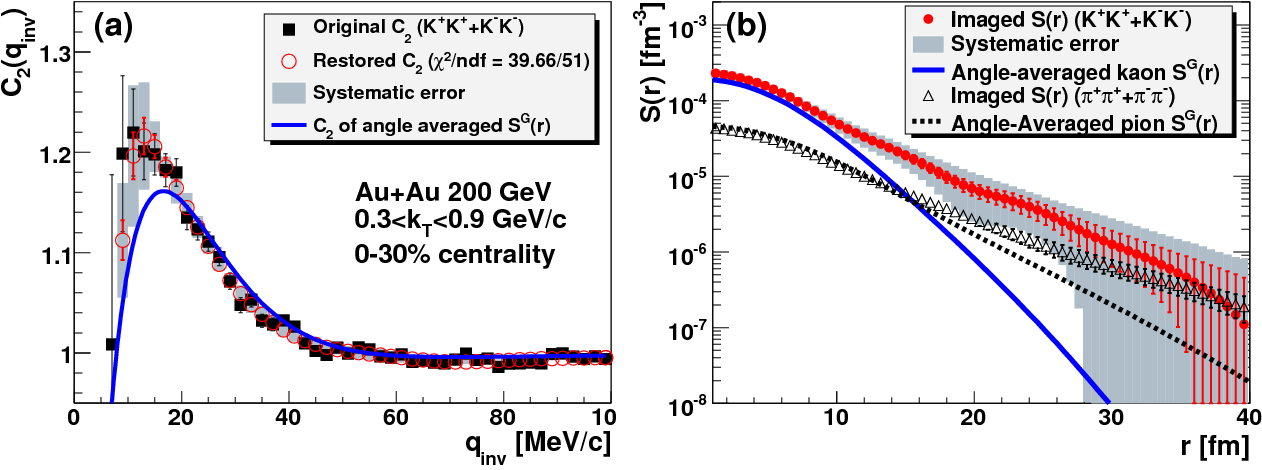}
\caption{On the top plot, correlation functions are shown.
Measurements and restored correlation functions are in
nice agreement. On the bottom plot imaged kaon
sources are shown. The deviation from the
Gaussian is larger than in case of pions.\label{f:Kaon_1Dimaging}}
\end{figure}

\section{Pion interferometry in p+p collisions}
\begin{figure}
\centering
\includegraphics[width=0.8\linewidth]{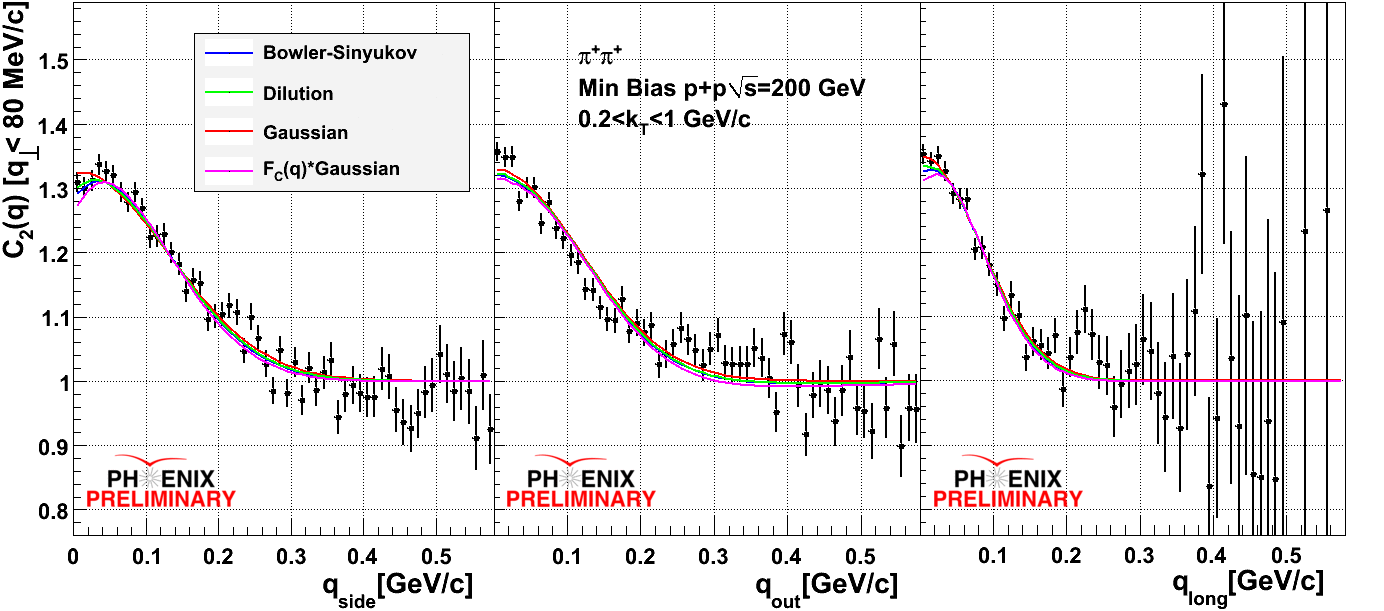}
\caption{Slices of three dimensional two-pion correlation functions measured in proton-proton collisions.}
\label{f:pp1d}
\end{figure}

\begin{figure}
\centering
\includegraphics[width=0.65\linewidth]{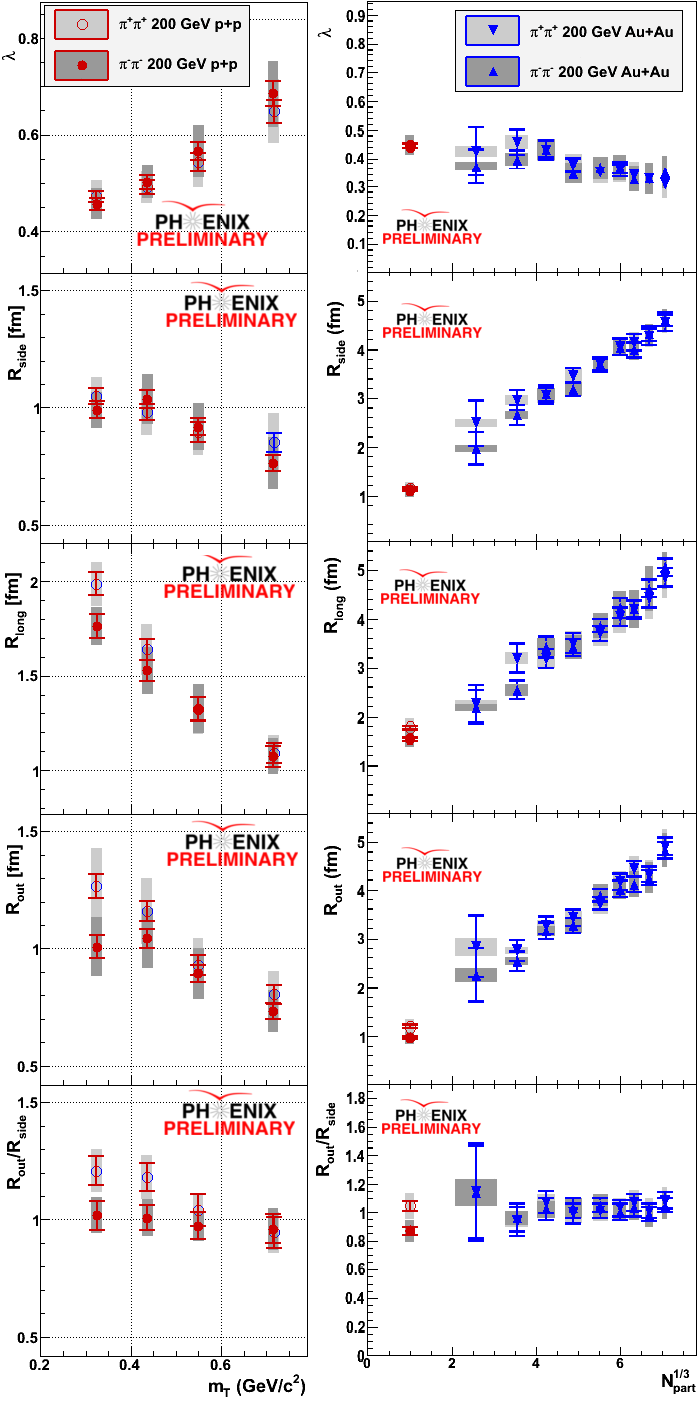}
\caption{The transverse mass (left) and number of participants (right) dependence of correlation parameters for pion pairs proton-proton collisions.
Final Au+Au HBT data is from ref.~\cite{Adler:2004rq}.}
\label{f:ppradii}
\end{figure}

The important measurement of Bose-Einstein correlations was extended to proton-proton systems
also. See more details of this analysis in ref.~\cite{Glenn:2009pf}.
PHENIX analyzed roughly 2.5 million like sign pion pairs from proton-proton
collisions of the 2004 and 2005 RHIC runs. Measurement techniques
are similar to those described in the previous section. One-dimensional
slices of the correlation function are shown in fig.~\ref{f:pp1d}, describable
by usual HBT techniques. Extracted source sizes are shown in fig.~\ref{f:ppradii}.
Usual $m_t$ behavior is observed, while the $N_{part}^{1/3}$ scaling curve of
Au+Au~\cite{Adler:2004rq} data is also in accordance with the new p+p results.

\section{Summary and conclusions}
We measured HBT correlation functions of charged kaon pairs in Au+Au collisions and of charged pion
pairs in p+p collisions. The 3D HBT radii are consistent for pions and kaons at the same number
of participants and transverse mass. The 1D emission source function for kaons extracted by imaging 
shows a non-Gaussian tail at distances greater than 10 fm. The preliminary analysis of pion HBT
correlations in p+p collisions can be analyzed via traditional 3D HBT methods. These correlation
radii are also consistent with extrapolations from earlier measurements.

\end{document}